# Neutrinos & Non-proliferation in Europe

Michel Cribier[*]
*APC, Paris*
*CEA/Saclay, DAPNIA/SPP*

The International Atomic Energy Agency (IAEA) is the United Nations agency in charge of the development of peaceful use of atomic energy. In particular IAEA is the verification authority of the Treaty on the Non-Proliferation of Nuclear Weapons (NPT). To do that jobs inspections of civil nuclear installations and related facilities under safeguards agreements are made in more than 140 states.

IAEA uses many different tools for these verifications, like neutron monitor, gamma spectroscopy, but also bookeeping of the isotopic composition at the fuel element level before and after their use in the nuclear power station. In particular it verifie that weapon-origin and other fissile materials that Russia and USA have released from their defense programmes are used for civil application.

The existence of an antineutrino signal sensitive to the power and to the isotopic composition of a reactor core, as first proposed by Mikaelian et al. [Mik77] and as demonstrated by the Bugey [Dec95] and Rovno experiments, [Kli94], could provide a means to address certain safeguards applications. Thus the IAEA recently ask members states to make a feasibility study to determine whether antineutrino detection methods might provide practical safeguards tools for selected applications. If this method proves to be useful, IAEA has the power to decide that any new nuclear power plants built has to include an antineutrino monitor.

Within the Double Chooz collaboration, an experiment [Las06] mainly devoted to study the fundamental properties of neutrinos, we thought that we were in a good position to evaluate the interest of using antineutrino detection to remotely monitor nuclear power station. This effort in Europe, supplemented by the US effort [Ber06], will constitute the basic answer to IAEA of the neutrino community.

---

[*] On behalf of a collective work by S. Cormon, M. Fallot, H. Faust, T. Lasserre, A. Letourneau, D. Lhuillier, V. Sinev from DAPNIA, Subatech and ILL.

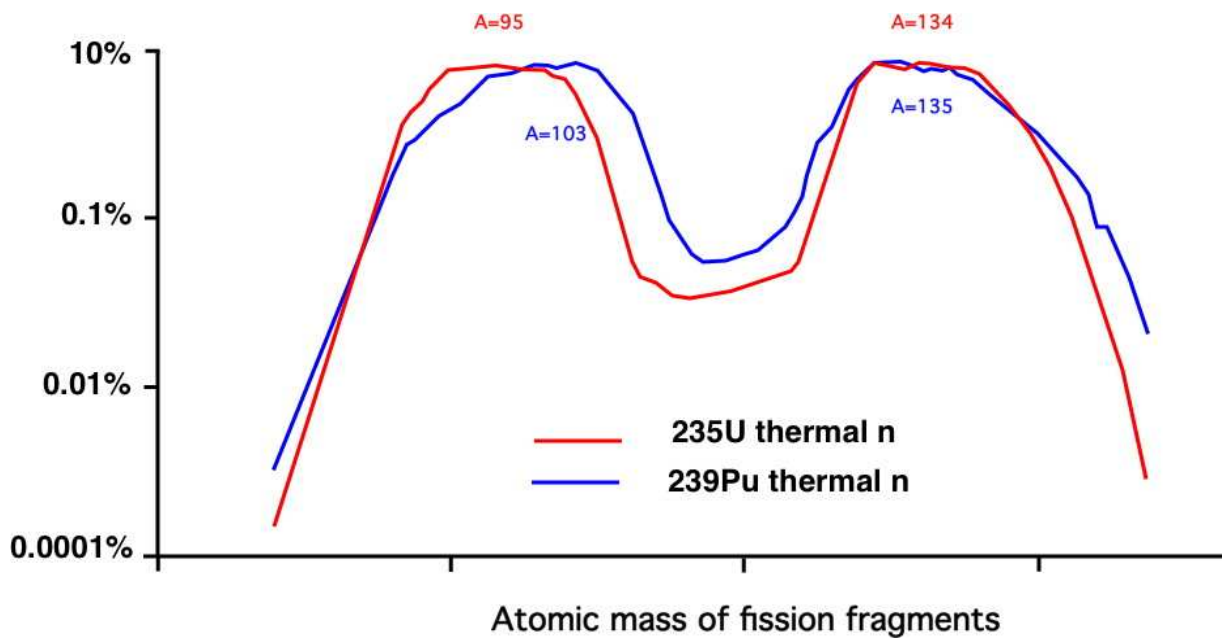

*Figure 1 : The statistical distribution of the fission products resulting from the fission of the most important fissile nuclei $^{235}$U and $^{239}$Pu shows two humps, one centered around masses 100 and the other one centered around 135. The low mass hump is at higher mass in $^{239}$Pu fission than in $^{235}$U, resulting in different nuclei and decays.*

The high penetration power of antineutrinos and the detection capability might provide a means to make remote, non-intrusive measurements of plutonium content in reactors [Ber02]. The antineutrino flux and energy spectrum depends upon the thermal power and on the fissile isotopic composition of the reactor fuel. Indeed, when a heavy nuclei (Uranium, Plutonium) experience a fission, it produce two unequal fission fragments (and a few free neutrons) ; the statistical distribution of the atomic masses is depicted in figure 1. All these nuclei immediately produced are extremely unstable - they are too rich in neutrons - and thus ß decay toward stable nuclei with an average of 6 ß decays. All these process involving several hundreds of unstable nuclei, with their excited states, makes very difficult to understand details of the physics, moreover, the most energetic antineutrinos, which are detected more easily, are produced in the very first decays, involving nuclei with typical lifetime smaller than a second.

|  | $^{235}$U | $^{239}$Pu |
|---|---|---|
| released energy per fission | 201.7 MeV | 210.0 MeV |
| Mean energy of ν | 2.94 MeV | 2.84 MeV |
| ν per fission > 1.8 MeV | 1.92 | 1.45 |
| average inter. cross section | $\approx 3.2\ 10^{-43}$ cm$^2$ | $\approx 2.76\ 10^{-43}$ cm$^2$ |

Based on predicted and observed ß spectra, the number of antineutrinos per fission from $^{239}$Pu is known to be less than the number from $^{235}$U, and the energy released bigger by 5%. Hence an hypothetical reactor able to use only $^{235}$U would induce in a detector an antineutrino signal 60% higher than the same reactor producing the same amount of energy but burning only $^{239}$Pu (see table). This offers a means to monitor changes in the relative amount of $^{235}$U and $^{239}$Pu in the core. If made in conjunction with accurate independent measurements of the thermal power (with the temperature and the flow rate of cooling water), antineutrino measurements might provide an estimate of the isotopic composition of the core, in particular its plutonium inventories. The shape of the antineutrino spectrum can provide additional information about core fissile isotopic composition.

Because the antineutrino signal from the reactor decreases as the square of the distance from the reactor to the detector a precise "remote" measurement is really only practical at distances of a few tens of meters if one is constrained to "small" detectors of the order of few cubic meter in size.

## Simulations

### MAGNITUDES OF SOME EFFECTS

In our group, the development of detailed simulations using professional reactor codes started (see below), but it seems wise to use less sophisticated methods in order to evaluate already, with some flexibility, the magnitude of some effects. To do that we started from the set of Bateman equations, as depicted graphicaly in figure 2, which discribed the evolution of fuel elements in a reactor. The gross simplification in such treatment is the use of average cross section, depending only on 3 groups (thermal neutron, resonance region, fast neutrons), and moreover the fact that the neutron flux is imposed and not calculated.

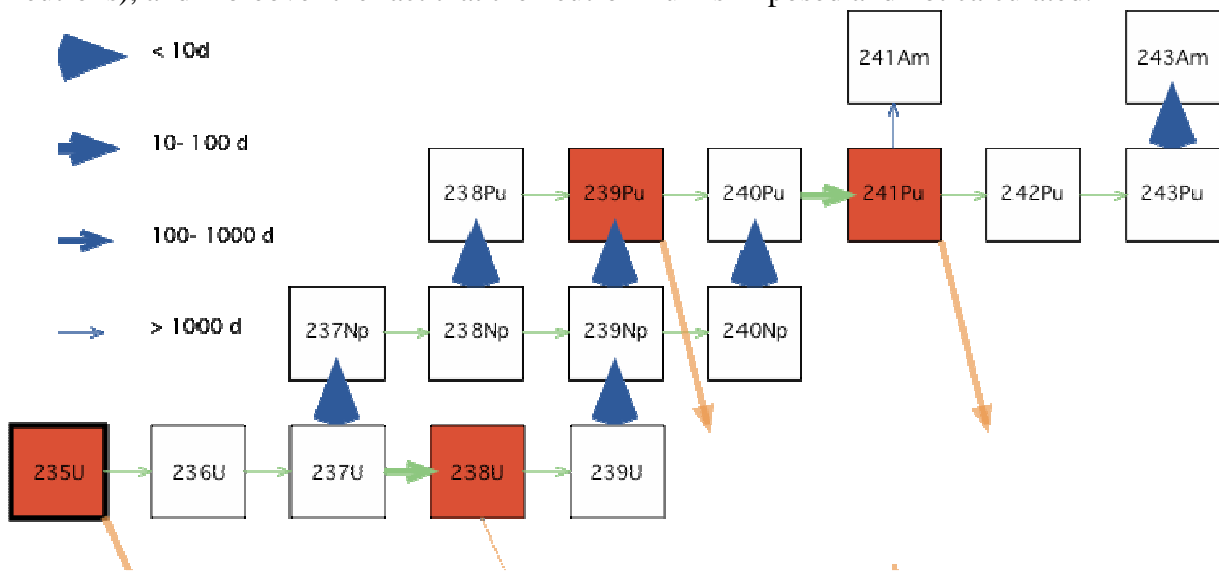

*Figure 2 : The Bateman equations are the set of differential equations which described all transformations of the nuclei submitted to a given neutron flux : capture of neutrons are responsible to move at Z constant (green arrow), ß-decay are responsible to increase the atomic mass by one unit (dark blue arrow), and fission destroy the heavy nuclei and produce energy (orange arrows).*

Given this we use for each isotope under consideration, the cross section for capture, fission, and also plug in the parameters of the decays. Then it is rather easy (and fast) to simulate the evolution of a given initial core composition ; in the same way, it is possible to « make a diversion » by manipulation the fuel composition at a choosen moment. As an example, the figure 3 show the evolution of a fresh core composed of Uranium enriched at 3.5 % in $^{235}$U : the build up of $^{239}$Pu and $^{241}$Pu is rather well reproduced.

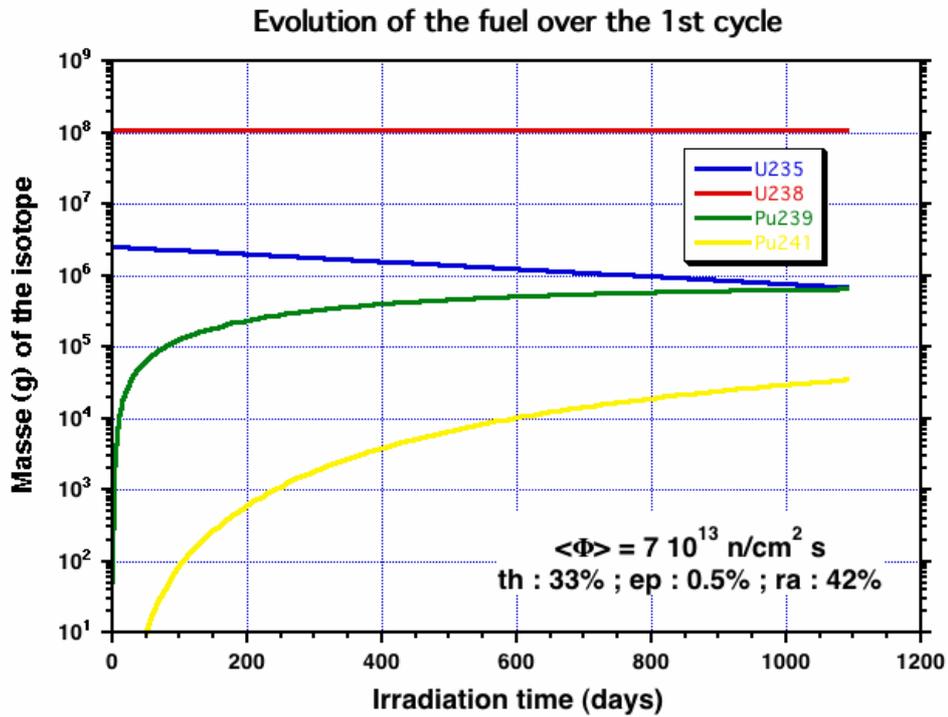

*Figure 3 : In a new reactor the initial fuel consist of enriched uranium rods, with an $^{235}U$ content typically at 3.5 %, the rest is $^{238}U$. As soon as the reactor is operating, reactions described by Bateman equations produce $^{239}Pu$ (and $^{241}Pu$), which then participate to the energy production, at the expense of $^{238}U$.*

Knowing the amount of fissions at a given time, it is straight forward to translate that in a given antineutrino flux using the parametrisation of [Hub04], and finally using the interaction cross section for inverse ß decay reaction, to produce the recorded signal in a given detector placed at a suitable location from the reactor under examination.

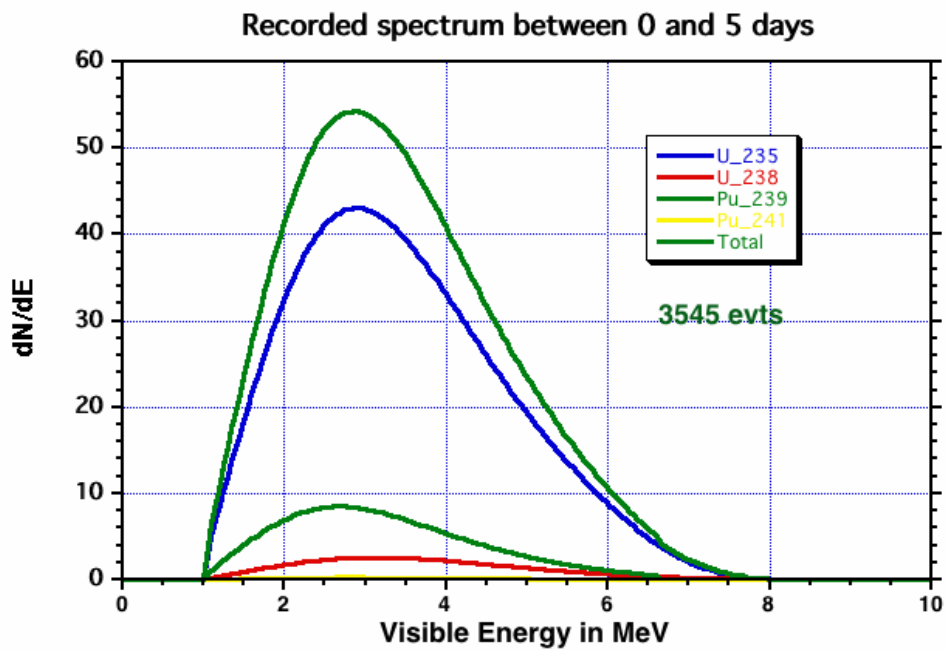

*Figure 4 : Positron spectrum recorded in an typical antineutrino detector (10 tons of target) placed at 150m of a nuclear reactor (1000 MW$_{el}$). Positrons results from the inverse ß-decay reaction used in the detection of anti-neutrino. The signal is the superposition of several component whose spectrum exhibit small but sizeable differences, especialy at high energy.*

As an example of this type of computation, we show in figure 5, the effect of the modification of fuel composition after 100 days : here the operator, clever enough, knows that he cannot merely remove Plutonium from the core without changing the thermal power which will be immediatly noticed. Hence he takes the precaution to add 28 kg of $^{235}$U at the same time where he remove 20 kg of $^{239}$Pu : although the thermal power is kept constant, the imprint on the antineutrino signal, although modest, is such that, after 10 days, there is an increase of more than 1 σ in the number of interactions recorded. Such a diversion is clearly impossible in PWR or BWR, but more easy in Candu-type reactor, and even more in a molten salt reactor.

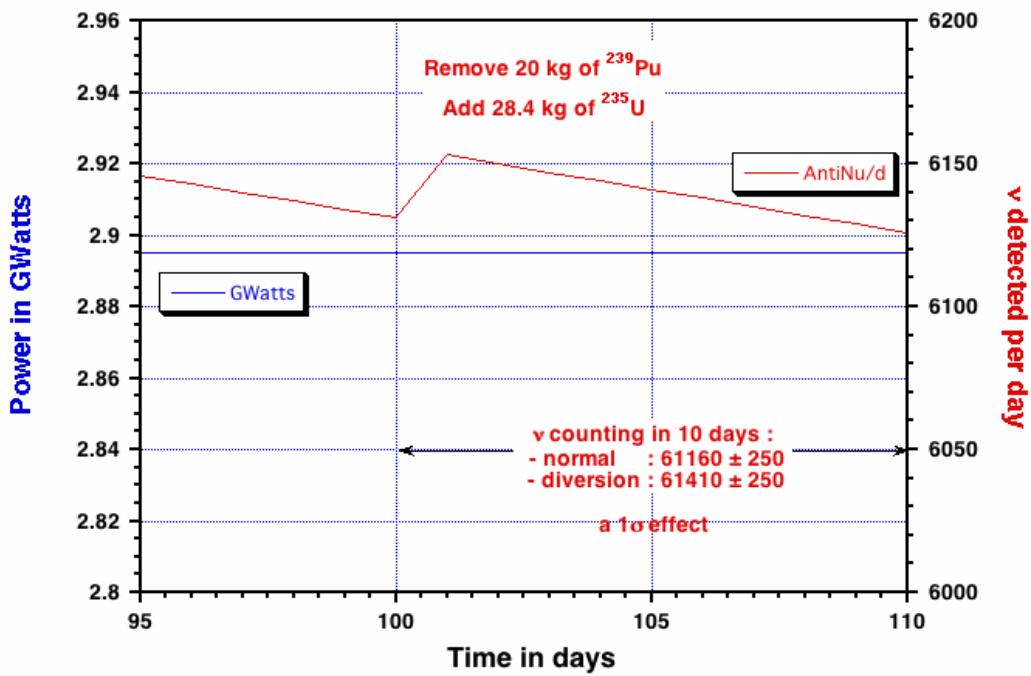

*Figure 5 : An hypothetical diversion scenario where an exchange of $^{239}$Pu with $^{235}$U is made such that the power does not change, but the antineutrino signal recorded by the monitor is slightly increased, giving some evidence of an abnormal operation.*

SIMULATIONS OF DIVERSION SCENARIOS

The IAEA recommends the study of specific safeguards scenarios. Among its concerns are the confirmation of the absence of unrecorded production of fissile material in declared reactors and the monitoring of the burn-up of a reactor core. The time required to manufacture an actual weapon estimated by the IAEA (conversion time), for plutonium in partially irradiated or spent fuel, lies between 1 and 3 months. The significant quantity of Pu is 8 kg, to be compared with the 3 tons of $^{235}$U contained in a Pressurized Water Reactor (PWR) of power 900MWe enriched to 3%. The small magnitude of the researched signal requires a carefull feasability study.

The proliferation scenarios of interest involve different kinds of nuclear power plants such as light water or heavy water reactors (PWR, BWR, Candu...), it has to include isotope production reactors of a few tens of MWth, and future reactors (e.g., PBMRs, Gen IV reactors, accelerator-driven sub-critical assemblies for transmutation, molten salt reactors). To perform these studies, core simulations with dedicated Monte-Carlo codes should be provided, coupled to the simulation of the evolution of the antineutrino flux and spectrum over time.

We started a simulation work using the widely used particle transport code MCNPX [Mcn05], coupled with an evolution code solving the Bateman equations for the fission products within a package called MURE (MCNP Utility for Reactor Evolution) [Mur05]. This package offers a set of tools, interfaced with MCNP or MCNPX, that allows to define easily the geometry of a reactor core. In the evolution part, it accesses, the set of evaluated nuclear data and cross sections. MURE is perfectly adapted to simulate the evolution with time of the composition of the fuel, taking into account the neutronics of a reactor core. We are adapting the evolution code to simulate the antineutrino spectrum and flux, using simple Fermi decay as starting point.

The extended MURE simulation will allows to perform sensitivity studies by varying the Pu content of the core in the relevant scenarios for IAEA. By varying the reactor power, the possibility to use antineutrinos for power monitoring can be evaluated.

Preliminary results show that nuclei with half-lives lower than 1s emit about 70% (50%) of the $^{235}$U( $^{239}$Pu) antineutrino spectrum above 6 MeV. The high energy part of the spectrum is the energy region where Pu and U spectra differ mostly. The influence of the ß decay of these nuclei on the antineutrino spectrum might be preponderant also in scenarios where rapid changes of the core composition are performed, e.g. in reactors such as Candu, refueled on line.

The appropriate starting point for this scenario is a representative PWR, like the Chooz reactors. For this reactor type, simulations of the evolution of the antineutrino flux and spectrum over time will be provided and compared to the accurate measurement provided by the near detector of Double Chooz. This should tell the precision on the fuel composition and of an independent thermal power measurements. An interesting point to study is at the time of the partial refuelling of the core, thanks to the fact that reactors like Chooz (N4-type) does not use MOX fuel.

Without any extra experimental effort, the near detector of the Double Chooz experiment will provide the most important dataset of anti neutrino detected ($5 \times 10^5$ ν per year) by a PWR. The precise neutrino energy spectrum recorded at a given time will be correlated to the fuel composition and to the thermal power provided by EDF. This valuable dataset will constitute an excellent experimental basis for the above feasibility studies of potential monitoring and for bench-marking fuel management codes ; it is expected that individual component due to fissile element ($^{235}$U, $^{239}$Pu) could be extracted with some modest precision and serve as a benchmark of this techniques.

To fulfil the goal of non-proliferation additional lab tests and theoretical calculations should also be performed to more precisely estimate the underlying neutrino spectra of plutonium and uranium fission products, especially at high energies. Contributions of decays to excited states of daughter nuclei are mandatory to reconstruct the shape of each spectrum. Following the conclusion of P. Huber and Th. Schwetz [Hub04] to achieve this goal a reduction of the present errors on the anti-neutrino fluxes of about a factor of three is necessary. We will see that such improvement needs an important effort.

**Experimental effort**

The precise measurement of β-decay spectra from fission products produced by the irradiation of a fissile target can be performed at the high flux reactor at Institut Laue

Langevin (ILL) in Grenoble, where similar studies performed in the past [Sch85] are the basis of the actual fluxes of antineutrinos used in these reactor neutrino experiment. The ILL reactor produces the highest neutron flux in the world : the fission rate of a fissile material target placed close to the reactor core is about $10^{12}$ per second. It is possible to choose different fissile elements as target in order to maximize the yield of the nucleus of interest. Using the LOHENGRIN recoil mass spectrometer [Loh04], measurement of individual β−spectra from short lived fission products are possible ; in the same irradiation channel, measurements of integral ß-spectrum with the Mini-INCA detectors [Mar06], could be envisaged to perform study on the evolution with time of the antineutrino energy spectrum of a nuclear power plant.

### EXPERIMENTS WITH LOHENGRIN

The LOHENGRIN recoil mass spectrometer offers the possibility to measure β-decays of individual fission products. The fissile target ($^{235}$U, $^{239}$Pu, $^{241}$Pu, …) is placed into a thermal neutron flux of $6.10^{14}$ n/cm$^2$/s, 50 cm from the fuel element. Recoil fission products are selected with a dipolar magnetic field followed by an electrostatic condenser. At the end the fragments could be implanted in a moving tape, and the measurement of subsequent β and γ-rays are recorded by a β-spectrometer (Si-detector) and Ge-clover detectors, respectively. Coincidences between these two quantities could also be made to reconstruct the decay scheme of the observed fission products or to select one fission product. Fragments with half-lives down to 2 μs can be measured, so that nuclei with large $Q_ß$ (above 4 MeV) can be measured.

The LOHENGRIN experimental objectives are to complete existing β-spectra of individual fission products [Ten89] with new measurements for the main contributors to the detected ν-spectra and to clarify experimental disagreements between previous measurements. This ambitious experimental programme is motivated by the fact - noted by C. Bemporad [Bem02] - that unknown decays contribute as much as 25% of the antineutrinos at energies > 4MeV. Folding the antineutrino energy spectrum over the detection cross-section for inverse beta decay enhances the contribution of the high energy antineutrinos to the total detected flux by a factor of about 10 for $E_\nu > 6$ MeV. The focus of these experiments will be on neutron rich nuclei with yields very different in $^{239}$Pu and $^{235}$U fission. In the list : $^{86}$Ge, $^{90-92}$Se, $^{94}$Br, $^{96-98}$Kr, $^{100}$Rb, $^{100-102}$Sr, $^{108-112}$Mo, $^{106-113}$Tc, $^{113-115}$Ru…contribute to the high energy part of the spectrum and have never been measured.

### IRRADIATION TESTS IN SUMMER 2005

A test-experiment has been performed during two weeks last in summer 2005. The isobaric chains A=90 and A=94 were studied where some isotopes possess a high $Q_ß$ energy, contributing significantly to the high energy part of the antineutrino spectra following $^{235}$U and $^{239}$Pu fissions and moreover produced with very different fission yields after $^{235}$U and $^{239}$Pu fission [Eng94]. The well-known nuclei, such as $^{90}$Br, will serve as a test of the experimental set-up, while the beta decay of more exotic nuclei such as $^{94}$Kr and $^{94}$Br will constitute a test case for how far one can reach in the very neutron rich region with this experimental device. The recorded data (figure 6) will validate the simulation described in the previous section, in particular the evolution over time of the isobaric chains beta decay spectra.

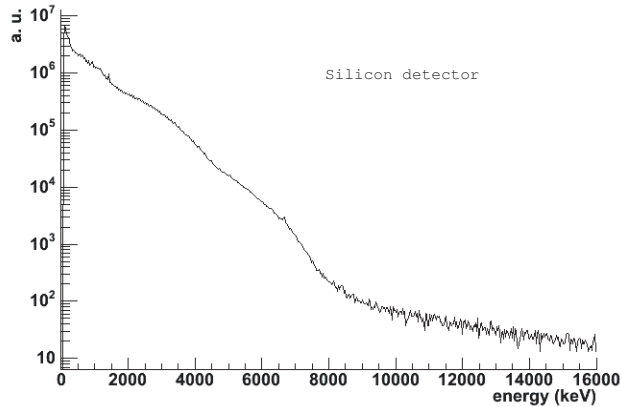

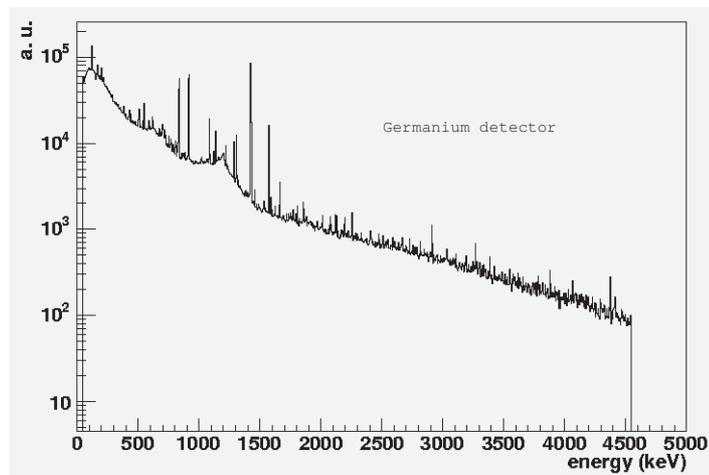

*Figures 6 : Beta energy spectrum (6a) recorded with the silicon detector corresponding to ß decay of fission products with mass A=94. The fission products arising from the LOHENGRIN spectrometer were implanted on a mylar tape of adjustable velocity in front of the silicon detector. The highest velocity was selected in order to enhance shorter-lived nuclei such as $^{94}Kr$ and $^{94}Br$. The gamma energy spectrum (6b) obtained with the germanium detector corresponding and to the same runs is displayed also.*

INTEGRAL ß SPECTRA MEASUREMENTS

In complement to individual studies on LOHENGRIN, more integral studies can be envisaged using the so called "Mini-INCA chamber" at ILL [Mar06] in return for adding a β-spectrometer (to be developed). The existing α- and γ-spectroscopy station is connected to the LOHENGRIN channel and offers the possibility to perform irradiations in a quasi thermal neutron flux up to 20 times the nominal value in a PWR. Moreover, the irradiation can be repeated as many time as needed. It offers then the unique possibility to characterize the evolution of the ß spectrum as a function of the irradiation time and the irradiation cooling. The expected modification of the β spectrum as a function of the irradiation time is connected to the transmutation induced by neutron capture of the fissile and fission fragment elements. It is thus related to the "natural" evolution of the spent-fuel in the reactor. The modification of the β spectrum as a function of the cooling time is connected to the decaying chain of the fission products and is then a means to select the emitted fragments by their livetime. This information is important because long-lived fission fragments accumulate in the core and after few days mainly contribute to the low energy part of the antineutrino-spectra.

Due to the mechanical transfer of the sample from the irradiation location to the measurement station an irreducible delay time of 30 mn is imposed leading to the loss of short-live fragments.

PROSPECT TO STUDY FISSION OF $^{238}$U

The integral beta decay spectrum arising from $^{238}$U fission has never been measured. All information relies on theoretical computations [Vog89]. Some experiments could be envisaged using few MeV neutron sources in Europe (Van de Graaf in Geel, SINQ in PSI, ALVARES or SAMES accelerators at Valduc, …). Here the total absence of experimental data on the ß emitted in the fission of $^{238}$U change the context of this measurement compared to the other isotopes. Indeed any integral measurements performed could be used to constraint the present theoretical estimations of the antineutrino flux produced in the fission of $^{238}$U. In any case it seems rather difficult to fulfil the goal of a determination of the isotopic content from antineutrinos measurements as long as in important part of the energy spectrum is so poorly known.

## Conclusions

After the preliminary studies, some thoughts can already be made. A realistic diversion (≈ 10 kg Pu) has an imprint in the antineutrino signal which is very small. The present knowledge on antineutrino spectrum emitted in fissions is not precise enough to allow a determination of the isotopic content in the core sensitive to such diversion.

On the other hand, the thermal power measurement is a less difficult job. Neutrinos sample the whole core, without attenuation, and would bring valuable information on the power with totally different systematics than present methods.

Even if its measurement is not dissuasive by itself, the operator cannot hide any stops or change of power, and in most case, such a record made with an external and independent device, virtually impossible to fake, will act as a strong constraint.

In spite of the uncertainty mentioned previously, we see that the most energetic part offers the best possibility to disentangle fission from $^{235}$U and $^{239}$Pu. The comparison between the cumulative numbers of antineutrinos as a function of antineutrino energy detected at low vs. high energy is an efficient observable to distinguish pure $^{235}$U and $^{239}$Pu.

IAEA seeks also monitoring large spent-fuel elements. For this application, the likelihood is that antineutrino detectors could only make measurements on large quantities of beta-emitters, e.g., several cores of spent fuel. In the time of the experiment the discharge of parts of the core will happen and the Double-Chooz experiment will quantify the sensitivity of such monitoring.

More generally the techniques developed for the detection of antineutrinos could be applied for the monitoring of nuclear activities at the level of a country. Hence a KamLAND type detector deeply submerged off the coast of the country, would offer the sensitivity to detect a new underground reactor located at several hundreds of kilometers. All these common efforts toward more reliable techniques, remotely operated detectors, not to mention undersea techniques will automatically benefit to both fields, safeguard and geo-neutrinos.

## References

[Bem02] Bemporad et al.,Rev. of Mod. Phys., Vol. 74, (2002).

[Ber02] A. Bernstein, Y. Wang, G. Gratta, and T. West, J. Appl. Phys. 91, 4672 (2002)

[Ber06] A. Bernstein, these proceeding

[Dec95] Y. Declais et al., Nucl. Phys. B434, 503 (1995)

[Eng94] T.R. England and B.F. Rider, ENDF-349, LA-UR-94-3106.

[Hub04] P. Huber, Th. Schwetz, Precision spectroscopy with reactor anti-neutrinos Phys.Rev. D70 (2004) 053011

[Kli94] Klimov et al., Atomic Energy, v.76-2, 123, (1994)

[Las06] T. Lasserre, these proceeding

[Loh04] ILL Instrument Review, 2004/2005.

[Mik77] Mikaelian L.A. Neutrino laboratory in the atomic plant, Proc. Int. Conference Neutrino-77, v. 2, p. 383-387

[Mar06] F. Marie, A. Letourneau et al., Nucl. Instr and Meth A556 (2006) 547.

[Mcn05] Monte Carlo N-Particle eXtended, LA-UR-05-2675, J.S.Hendricks et al.

[Mur05] MURE : MCNP Utility for Reactor Evolution -Description of the methods, first applications and results. MÃl'plan O., Nuttin A., Laulan O., David S., Michel-Sendis F. et al. In Proceedings of the ENC 2005 (CD-Rom) (2005) 1-7.

[Sch85] K. Schreckenbach, G. Colvin, W. Gelletly, F.v. Feilitzch, Phys. Lett. B160 (1985) 325

[Ten89] O. Tengblad et al., Nuclear Physics A 503 (1989) 136-160.

[Vog89] P. Vogel and J. Engel, Phys. Rev. D39, 3378 (1989)